\documentclass[aps, prd, twocolumn, amsmath, floats,floatfix, superscriptaddress, nofootinbib, showkeys,
showpacs]{revtex4}
\usepackage{amssymb}
\usepackage{amsmath}
\usepackage{verbatim}
\usepackage{mathrsfs}
\usepackage{amsfonts}
\usepackage{latexsym}
\usepackage{epsfig}
\usepackage{color}
\usepackage{graphicx,subfigure}
\usepackage{units}
\usepackage{envmath}
\usepackage{natbib}
\usepackage{ctable}
\usepackage{soul}
\begin{document}
\definecolor{orange}{rgb}{0.9,0.45,0}
\newcommand{\re}{\mbox{Re}}
\newcommand{\im}{\mbox{Im}}
\newcommand{\tf}[1]{\textcolor{red}{TF: #1}}
\newcommand{\nsg}[1]{\textcolor{cyan}{#1}}
\newcommand{\saeed}[1]{\textcolor{blue}{SF: #1}}
\newcommand{\fdg}[1]{\textcolor{orange}{FDG: #1}}
\newcommand{\jc}[1]{\textcolor{magenta}{JC: #1}}
\def\CovDev{D}
\def\Res{{\mathcal R}}
\def\Gammaflat{\hat \Gamma}
\def\metricflat{\hat \gamma}
\def\Dflat{\hat {\mathcal D}}
\def\part_n{\partial_\perp}
%
\def\Lie{\mathcal{L}}
\def\A{\mathcal{X}}
\def\Aphi{\A_{\phi}}
\def\hAphi{\hat{\A}_{\phi}}
\def\E{\mathcal{E}}
\def\Ham{\mathcal{H}}
\def\M{\mathcal{M}}
\def\R{\mathcal{R}}
\def\p{\partial}
\def\hg{\hat{\gamma}}
\def\hA{\hat{A}}
\def\hD{\hat{D}}
\def\hE{\hat{E}}
\def\hR{\hat{R}}
\def\hcA{\hat{\mathcal{A}}}
\def\hDelt{\hat{\triangle}}
\def\na{\nabla}
\def\dif{{\rm{d}}}
\def\non{\nonumber}
\newcommand{\erf}{\textrm{erf}}
\renewcommand{\t}{\times}
\long\def\symbolfootnote[#1]#2{\begingroup%
\def\thefootnote{\fnsymbol{footnote}}\footnote[#1]{#2}\endgroup}
\title{Primordial Black Hole Merger Rate in Ellipsoidal-Collapse Dark Matter Halo Models}

\author{Saeed Fakhry}
\email{s_fakhry@sbu.ac.ir}
\affiliation{Department of Physics, Shahid Beheshti University,
Evin, Tehran 19839, Iran}

\author{Javad T. Firouzjaee}
\email{firouzjaee@kntu.ac.ir}
\affiliation{Department of Physics, K.N. Toosi University of
Technology, P.O. Box 15875-4416, Tehran, Iran}
\affiliation{School of Physics, Institute for Research in Fundamental
             Sciences (IPM), P.O. Box 19395-5531, Tehran, Iran}

\author{Mehrdad Farhoudi}
\email{m-farhoudi@sbu.ac.ir}
\affiliation{Department of Physics, Shahid Beheshti University,
Evin, Tehran 19839, Iran}

\date{May 9, 2021}
\begin{abstract}
\noindent
 We have studied the merger rate of primordial black holes (PBHs)
in the ellipsoidal-collapse model of halo to explain the dark
matter abundance by the PBH merger estimated from the
gravitational waves detections via the Advanced LIGO (aLIGO)
detectors. We have indicated that the PBH merger rate within each
halo for the ellipsoidal models is more significant than for the
spherical models. We have specified that the PBH merger rate per
unit time and per unit volume for the ellipsoidal-collapse halo
models is about one order of magnitude higher than the
corresponding spherical models. Moreover, we have calculated the
evolution of the PBH total merger rate as a function of redshift.
The results indicate that the evolution for the ellipsoidal halo
models is more sensitive than spherical halo models, as expected
from the models. Finally, we have presented a constraint on the
PBH abundance within the context of ellipsoidal and spherical
models. By comparing the results with the aLIGO mergers during the
third observing run (O3), we have shown that the merger rate in
the ellipsoidal-collapse halo models falls within the aLIGO
window, while the same result is~not valid for the
spherical-collapse ones. Furthermore, we have compared the total
merger rate of PBHs in terms of their fraction in the
ellipsoidal-collapse halo models for several masses of PBHs. The
results suggest that the total merger rate of PBHs changes
inversely with their masses. We have also estimated the relation
between the fraction of PBHs and their masses in the
ellipsoidal-collapse halo model and have shown it for a narrow
mass distribution of PBHs. The outcome shows that the constraint
inferred from the PBH merger rate for the ellipsoidal-collapse
halo models can be potentially stronger than the corresponding
result obtained for the spherical-collapse ones.
\end{abstract}


\maketitle

\vspace{8cm}

\section{Introduction}

The detection of binary black holes with the
LIGO~\cite{Abbott:2016blz,Abbott:2016nmj,TheLIGOScientific:2017qsa,Abbott:2020khf,Abbott:2020tfl}
has opened a new epoch in probing the nature and behavior of
compact objects in the cosmos. In the past years, the
gravitational wave detectors have directly confirmed the existence
of black holes~\cite{Abbott:2016nmj}, and have provided powerful
tests of general relativity~\cite{TheLIGOScientific:2016src}.
These detectors are also guided in the era of multi-messenger
astronomy~\cite{GBM:2017lvd}. While the gravitational wave
observatories are continued to probe the population of black
holes, another significant discovery arises as to whether mergers
may provide direct evidence for the existence of PBHs.

It is widely accepted that the astrophysical objects are
originated from the early universe quantum fluctuations which
became classical as were stretched to super-horizon scales in an
exponentially expanding period. If the density perturbations of
these fluctuations exceed some threshold value, the PBHs might
form. Since passing by the threshold value is the critical point
of formation, many numerical investigations have been done to
study the threshold value for the density perturbations, see e.g.,
Refs.~\cite{Carr:1975qj,Niemeyer:1999ak,Musco:2004ak,Young:2,Shibata:1999zs,new-thre,musco,bloom,allah}.
Besides PBHs, there might be other formation channels for black
holes with the non-stellar beginning such as the gravitational
collapse in dark matter
candidates~\cite{Khlopov:2008qy,Belotsky:2014kca,Shandera:2018xkn},
or the collapse of other compact objects due to new
physics~\cite{deLavallaz:2010wp}. During the last two decades,
many works have been done in the subject of PBHs and related area,
see e.g., Ref.~\cite{Carr:2020xqk} and references therein.

Since the massive PBHs interact only via gravitation, and since a
large set of black holes behaves as perfect fluids on sufficiently
large scales, the PBHs are a natural nominee for dark matter.
Nowadays, though the existence of PBHs is yet neither proven nor
refuted, the very observational limits on its abundance represent
themselves a powerful and unique probe of the early universe at
small scales, which is difficult to probe with any other
method~\cite{Lehmann:2018ejc, Carr:2017jsz, Carr:2020gox}. Besides
other observational constraints such as Icarus
\cite{Carr:2020gox}, one of the (serious) bounds on the abundance
of PBHs in the mass range around $10-30~M_{\odot}$ can be obtained
from the LIGO observations that are assumed in involving the
merging of PBHs pairs. Shortly after the first observation of a
binary black hole merger, in Refs.~\cite{bird, Sasaki:2016jop}, it
has been stated that the merger rate by the LIGO discovery is
potentially consistent with a mass fraction of PBHs accounting for
the total of dark matter, assuming that those two black holes
involved had a primordial origin and LIGO had detected dark
matter.

Since the PBHs merge in the halo and consist of a fraction of dark
matter, the halo mass function can affect the merger rate of PBHs.
In addition, the concentration parameter changes the relative
velocity distribution of PBHs within each halo, which determines
their merger rate within each halo.

On the other hand, there are different types of halo collapse
models. The analytically simple model is the spherical-collapse
which has been found to over-predict the abundance of small halos
and under-predict for the massive ones. This issue is because the
halo collapses are generally triaxial rather than spherical. The
Sheth-Mo-Tormen model~\cite{Sheth:1999mn,st} uses the
ellipsoidal-collapse model and the obtained fitting functions
provide a closer match to the unconditional halo mass function in
N-body simulations. Furthermore, the ellipsoidal-collapse model
has its mass-concentration relation which gives deep insight into
the formation and structure of halos~\cite{afshordi}.

In this work, we propose to use the ellipsoidal-collapse model to
calculate the merger rate of PBHs, which are in the dark matter
halo. In this respect, the outline of the work is as follows. In
Sec.~\ref{sec. ii}, we introduce the dark matter halo model and
its concentration. Moreover, we discuss the halo mass function for
both the spherical- and ellipsoidal-collapse models. Then, in
Sec.~\ref{sec. iii}, we calculate the PBH merger rate in the
ellipsoidal-collapse model and compare it with the corresponding
results of the spherical-collapse model. Furthermore, we discuss
the constraints arising from the merger rate of PBHs in the
ellipsoidal-collapse model. Finally, we scrutinize the results and
summarize the findings in Sec.~\ref{sec. iv}.

\section{Halo Models}\label{sec. ii}
\subsection{Halo Density Profile} \label{subsec:hdp}
Dark matter halos are considered as nonlinear cosmological
structures whose masses can be modeled as a function of radius
called density profile. In recent years, the analytical models and
numerical simulations have provided a clearer picture of the
properties and behavior of these structures. In addition, many
studies have been performed to extract a convenient density
profile that can provide an acceptable description for the
observational
data~\cite{dehnen,Hernquist:1990be,Jaffe:1983iv,deZeeuw:1985sk,Plummer:1911zza,Navarro:1995iw,ein}.

Let us mention two of the most commonly used density profiles as
follows. The first one is the Navarro-Frenk-White ({\bf NFW})
profile, which is extracted from the $N$-body simulations and is
well compatible with most of the rotation curve
data~\cite{Navarro:1995iw}. The relation of this density profile
is
\begin{equation}\label{1a}
    \rho(r)=\frac{\rho_{\rm s}}{r/r_{s}(1+r/r_{s})^{2}}.
\end{equation}
The other one, which is derived from analytical models, is the
Einasto profile that is also well consistent with the
observational data~\cite{ein}, and has the form
\begin{equation}\label{2}
 \rho(r)=\rho_{\rm s} \exp \{-\frac{2}{\alpha}\left[\left(\frac{r}{r_{s}}\right)^{\alpha}-1\right]\}.
\end{equation}
In these relations, $\rho_{s}$ and $r_{s}$ are the scaled
parameters that vary from halo to halo, and $\alpha$ is the shape
parameter for the Einasto profile. It should be noted that for
both of the above forms, one has
\begin{equation}
    \frac{d \ln \rho(r)}{d \ln r} = -2 \hspace*{1cm} {\rm for} \hspace*{0.2cm} r/r_{\rm s} = 1,
\end{equation}
i.e. the logarithmic slope of the density distribution is $-2$.

On the other hand, the halo density profile can be described in
terms of two parameters, namely the mass and concentration. The
halo concentration describes the central density of halos and is
defined as
\begin{equation}\label{conc}
C \equiv \frac{r_{\rm vir}}{r_{\rm s}},
\end{equation}
where $r_{\rm vir}$ is a viral radius considered as a radius
within which the average halo density reaches $200$ to $500$ times
the critical density of the universe. Also, the $N$-body
simulations show that the concentration parameter is a decreasing
function of the halo mass and is a redshift dependent function at
fixed mass~\cite{afshordi,prada,Dutton:2014xda,ludlow}. We will
discuss about the mass distribution of dark matter halos in the
next subsection.

\subsection{Halo Mass Function}
The existence of dark matter halos provides a convenient and
fundamental framework to study nonlinear gravitational collapse in
the universe. Hence, having a proper statistical view of the mass
distribution of these halos can improve our understanding of the
physics governing those. With this argument, a function called the
halo mass function has been proposed, which describes the mass
distribution of these halos within a given volume. In other words,
the halo mass function describes the mass of those structures
whose overdensities exceed the threshold overdensity, separate
from the expansion of the universe, and collapse. Furthermore, in
the standard cosmology, one may define a linear quantity called
density contrast as
$\delta(x)\equiv[\rho(x)-\bar{\rho}]/\bar{\rho}$, where $\rho(x)$
is the local density at any point $x$ and $\bar{\rho}$ is the mean
background energy density. As noted earlier, this quantity may
grow to the critical point while the universe expands, exceeds
linear regimes, and enters into nonlinear regimes. This situation
occurs when the overdensities separate from the expansion of the
universe, enter the turnaround phase and collapse. That is, the
structures are formed at this stage. For an Einstein-de Sitter
universe and a spherical-collapse halo model, the threshold
overdensity has been calculated to be $\delta_{\rm sc} =1.686$.
This threshold depends only on the redshift value and is
independent of all local quantities such as mass and
radius~\cite{Lukic:2007fc}. Hence, it can be considered as a fixed
threshold in a narrow redshift range. However, this threshold
varies for ellipsoidal-collapse halo models, which we will discuss
later.

In Ref.~\cite{Jenkins:2000bv}, in order to specify various fits
for dark matter halos, an appropriate definition of the
differential mass function has been presented as
\begin{equation} \label{mf}
    \frac{dn}{dM}=g(\sigma)\frac{\rho_{\rm m}}{M}\frac{d\ln(\sigma^{-1})}{dM},
\end{equation}
where $n(M)$ is the number density of halos with mass $M$,
$\rho_{m}$ is the cosmological matter density, and $g(\sigma)$
depends on the geometry of overdensities at the collapse time. The
function $\sigma(M,z)$ is the linear root mean square fluctuation
of overdensities on mass $M$ and redshift $z$, which is defined as
\begin{equation}
    \sigma^{2}(M,z) \equiv\frac{1}{2\pi^{2}}\int_{0}^{\infty}P(k,z)W^{2}(k,M)k^{2}dk.
\end{equation}
In this relation, $W(k, M)$ is the Fourier spectrum of the top-hat
filter which depends on mass $M$ and wavenumber $k$, and $P(k,z)$
is the power spectrum of the fluctuations.

There is a wide range of studies that have been conducted to
extrapolate the halo mass function based on analytical approaches
and numerical simulations. The purpose of these studies is to
provide the best fit for the cosmic observations. The first model
for the dark matter halo mass function, assuming a homogeneous and
isotropic collapse, was presented by Press and Schechter ({\bf
P-S})~\cite{ps} as
\begin{equation}\label{mf1}
g_{\rm ps}(\sigma) = \sqrt{\frac{2}{\pi}}\frac{\delta_{\rm
sc}}{\sigma}\exp\left(-\frac{\delta_{\rm
sc}^{2}}{2\sigma^{2}}\right),
\end{equation}
which is called the P-S mass function. This formalism is based on
the assumption that every astrophysical or cosmological object is
formed via a gravitational collapse of overdensities. Moreover,
although the final collapse is a nonlinear process, it is assumed
that, in the early universe, the density fluctuations had been
very small and resulted in a linear approximation. As is clear
from relation~(\ref{mf1}), at a fixed redshift, the mass function
depends only on the mass of halos via $\sigma(M)$, and it is
expected that no significant change can be observed. This mass
function has been proposed as the simplest model for the formation
of dark matter halos, i.e. a spherical-collapse model, and, in
many cases, is consistent with the observational data.
Nevertheless, it quantitatively deviates from the numerical
results at some mass limits~\cite{Jenkins:2000bv}. Therefore, some
improvements have been made to address this issue. One of the most
successful improvements was provided by Sheth and Tormen
{\bf(S-T)}, which is based on a more realistic model and fits
simulation results better~\cite{Sheth:1999mn,st}. Their formalism
was based on an ellipsoidal-collapse model with dynamical
threshold density fluctuations, in contrast to an almost global
threshold in the P-S model.

As mentioned earlier, the threshold overdensity for
spherical-collapses, $\delta_{\rm sc}$, has been introduced as a
global value. It means that, in about certain redshifts, all
structures with overdensities higher than such a threshold can
collapse. S-T have proposed the idea that dynamically considering
the threshold overdensity for the ellipsoidal-collapses,
$\delta_{\rm ec}$, can provide a more realistic picture of the
halo mass function. With this assumption and considering
prolateness to be zero~\cite{st}, they have extracted this
quantity as
\begin{equation}
    \delta_{\rm ec}(\nu)\approx\delta_{\rm sc}(1+\gamma\,\nu^{-2\beta}),
\end{equation}
with $\gamma = 0.47$, $\beta = 0.615$ and $\nu\equiv\delta_{\rm
sc}/\sigma(M)$. It is clear that this quantity not only implicitly
depends on the redshift, but also on the mass of the structure,
and is called the moving barrier. With this assumption, one can
find the halo mass function for the ellipsoidal-collapse, which is
also called the S-T mass function, to be
\begin{equation}\label{mf2}
 g_{\rm st}(\sigma) = F \sqrt{\frac{2a}{\pi}}\frac{\delta_{\rm sc}}
 {\sigma}\exp\left(-\frac{a\delta_{\rm sc}^{2}}{2\sigma^{2}}\right)
 \left[1+\left(\frac{\sigma^{2}}{a\delta_{\rm sc}^{2}}\right)^{p}\right],
\end{equation}
with $F = 0.3222$, $a = 0.707$ and $p = 0.3$. This mass function is
expected to be more sensitive than the P-S mass function with
redshift changes. Thus, we now have all the tools that one needs
to study PBHs in dark matter halos. In this regard, in the next
section, we will talk about the probability of encountering PBHs,
their binary formation, and their merger rate within a certain
volume and time interval.

\section{Primordial Black Hole Merger Rate}\label{sec. iii}
\subsection{Merger Rate Within Each Halo}
The PBHs are a special type of black holes that are formed in the
early universe due to the direct collapse of density fluctuations
or equivalently curvature perturbations. The PBHs were~not only
able to form binaries when the universe had been dominated by
radiation but also could encounter each other in the late time
universe due to their random distribution.

More important is that these types of black holes have been
considered as candidates for dark matter for over $40$ years. In
this regard, it is believed that the gravitational wave events
recorded by the LIGO detectors can be described by the PBH
scenario at typical mass $30~M_{\odot}$, if these black holes
could be considered as a component of dark matter. Hence, we have
focused on the stellar-mass PBHs that may reside in the galactic
halos and have potentially been proposed as a dark matter
candidate. Indeed, the recent detections of stellar-mass black
hole mergers have drawn much attention to study their origins, and
the PBH scenario is one of the plausible theories.

The presence of PBHs with random distributions in dark matter
halos gives those a chance to form binaries via the close
encounter and emitting gravitational waves. In particular, the
numerical simulations \cite{Moore:1999nt} and analytical
approaches \cite{Kamionkowski:2008vw} to investigate the subhalos
resided in the larger halos show that the dark matter in the
subhalos has lower velocity dispersion and higher density than the
host halos \cite{bird}. Hence, the subhalos are likely to have the
largest contribution to the formation of PBH binaries. Thus, the
probability of PBH merger in the subhalos is more significant than
in the host halos~\cite{bird}.

Suppose two PBHs with masses $m_{i}$ and $m_{j}$ and relative
velocity at the large separation $v_{\rm rel}=|v_{i}-v_{j}|$
accidentally encounter each other in a dark matter halo. At the
closest physical separation (i.e., at periastron), due to the
maximum scattering amplitude, a significant gravitational
radiation occurs. Regarding the emission of the gravitational
radiation $E_{\rm rad}$, the Keplerian mechanics implies that the
time-averaged radiated energy to be \cite{peter:1964}
\begin{equation}
\langle\frac{dE_{\rm rad}}{dt}\rangle = -\frac{32}{5}\frac{G^{4}
m_{i}^{2}m_{j}^{2}(m_{i}+m_{j})}{c^{5}a^{3/2}r_{\rm
p}^{7/2}(1+e)^{7/2}}
\left(1+\frac{73}{24}e^{2}+\frac{37}{96}e^{4}\right),
\end{equation}
where $G$ is the gravitational constant, $c$ is the velocity of
light, $a$ is the semi-major axis of the orbit, $e$ is the
eccentricity of the orbit and $r_{\rm p}=a(1-e)$ is the
periastron. As the head-on collision rarely happens, it can be
assumed that near the periastron, due to the emission of the
maximum gravitational radiation, the trajectory is an ellipse with
a maximum eccentricity (i.e. $e=1$). Under these assumptions,
after one orbital period $T=2\pi\sqrt{a^{3}/[G(m_{i}+m_{j})]}$,
the radiated gravitational energy is
\begin{equation}
\Delta E_{\rm rad} = \frac{85\pi
G^{7/2}m_{i}^{2}m_{j}^{2}\sqrt{m_{i}+m_{j}}}{12\sqrt{2}c^{5}r_{\rm
p}^{7/2}}.
\end{equation}
If $\Delta E_{\rm rad}>E_{\rm kin}$, where $E_{\rm kin} = \mu
v_{\rm rel}^{2}/2$ is the kinetic energy and $\mu =
m_{i}m_{j}/(m_{i}+m_{j})$ is the reduced mass, then the PBHs will
form binary systems. This condition provides a maximum for the
periastron as
\begin{equation}\label{0}
    r_{\rm p, max} = \left[\frac{85 \pi\sqrt{2}G^{7/2}m_{i}m_{j}(m_{i}+m_{j})^{3/2}}{12c^{5}v_{\rm rel}^{2}}\right]^{2/7},
\end{equation}
which means that $r_{\rm p} < r_{\rm p, max}$ leads to the binary
formation. In addition, in the Newtonian limit, the impact
parameter $b$ is calculated as a function of $r_{\rm p}$
\cite{Sasaki:2018dmp}, namely
\begin{equation}
b^{2}(r_{\rm p})=\frac{2G(m_{i}+m_{j})r_{\rm p}}{v_{\rm rel}^{2}}+r_{\rm p}^{2}.
\end{equation}
In other words, if the value of the impact parameter for the
encounter is less than $b(r_{\rm p,max})$, one can expect to form
a PBH binary. Therefore, the cross-section for the binary
formation is obtained to be~\cite{quinlan,Mouri:2002mc}
\begin{align}\label{1}
    \xi(m_{i}, m_{j}, v_{\rm rel}) = \pi b^{2}(r_{\rm p, max})\simeq \frac{2\pi G(m_{i}+m_{j}) r_{\rm p, max}}{v_{\rm rel}^{2}},
\end{align}
where the strong limit of the gravitational focusing, i.e. $r_{\rm
p} \ll b$, has been applied.

Our focus is on the merger rate of the PBHs that are compatible
with the LIGO sensitivity,
i.e.\,$\sim\left(30~M_{\odot}-30~M_{\odot}\right)$ events in the
galactic halos. Accordingly, we have normalized those masses to
$30~M_{\odot}$ with their relative velocities as the average
velocities of dark matter halos, i.e. $200~\rm km/s$.

By inserting Eq.~(\ref{0}) into Eq.~(\ref{1}) and assuming
$m_{i}=m_{j}=M_{\rm pbh}$ and $v_{\rm rel}=v_{\rm pbh}$, one can
reach an explicit form of the cross-section related to the
normalized mass and velocity of the PBHs as
\begin{multline}
\xi \simeq 4\pi \left(\frac{85\pi}{3}\right)^{2/7}\left(\frac{M_{\rm pbh}^{2}G^{2}}{c^{10/7}v_{\rm{pbh}}^{18/7}}\right) \\
\simeq 1.37 \times 10^{-14}\left(\frac{M_{\rm pbh}}{30M_{\odot}}\right)^{2}\left(\frac{v_{\rm pbh}}{200
\rm km/s}\right)^{-18/7} \rm in\hspace{0.3cm}(pc)^{2}.
\end{multline}
With these considerations, the merger rate of PBHs within each
halo can be calculated via the formula
\cite{bird,Nishikawa:2017chy}
\begin{equation}\label{13}
 \Phi = 2\pi\int_{0}^{r_{\rm vir}} r^{2}\left(\frac{\rho_{\rm halo}(r)}{M_{\rm pbh}}\right)^{2}\langle \xi v_{\rm pbh}\rangle dr,
\end{equation}
where $\rho_{\rm halo}(r)$ is the halo density profile that can be
considered to be the NFW or the Einasto density profile, and
$\langle\xi v_{\rm pbh}\rangle$ represents an average over the
relative velocity distribution of PBHs in the galactic halos.

Moreover, the mass located within the virial radius of the halo,
the virialized mass, can be found by
\begin{equation}\label{2a}
    M_{\rm vir}=\int_{0}^{r_{\rm vir}}4\pi r^{2}\rho(r) dr.
\end{equation}
By inserting relation (\ref{1a}) into relation (\ref{2a}) and
integrating, one can find the virialized mass for the NFW density
profile as
\begin{equation}\label{15}
M_{\rm vir\,(NFW)}= 4\pi \rho_{\rm s} r_{\rm s}^{3}\left(\ln(1+C)-\frac{C}{1+C}\right).
\end{equation}
Similarly, by considering relation (\ref{2}), the virialized mass
for the Einasto density profile~\cite{dan,retana} can be obtained
as
\begin{equation}\label{16}
M_{\rm vir\,(Ein)}= 4\pi \rho_{\rm s} r_{\rm s}^{3}l(C,\alpha).
\end{equation}
In this relation, $l(C,\alpha)$ is a function of concentration
and shape parameter and has the following form
\begin{equation*}
l(C,\alpha)=\frac{\exp(2/\alpha)}{\alpha}\left(\frac{\alpha}{2}\right)^{3/\alpha}\Gamma(\frac{3}{\alpha},\frac{2}{\alpha} C^{\alpha}),
\end{equation*}
where $\Gamma(x, y)=\int_{0}^{y}t^{x-1}e^{-t}dt$ is the incomplete Gamma function.

\begin{figure}[t!]
\begin{minipage}{1\linewidth}
\includegraphics[width=1\textwidth]{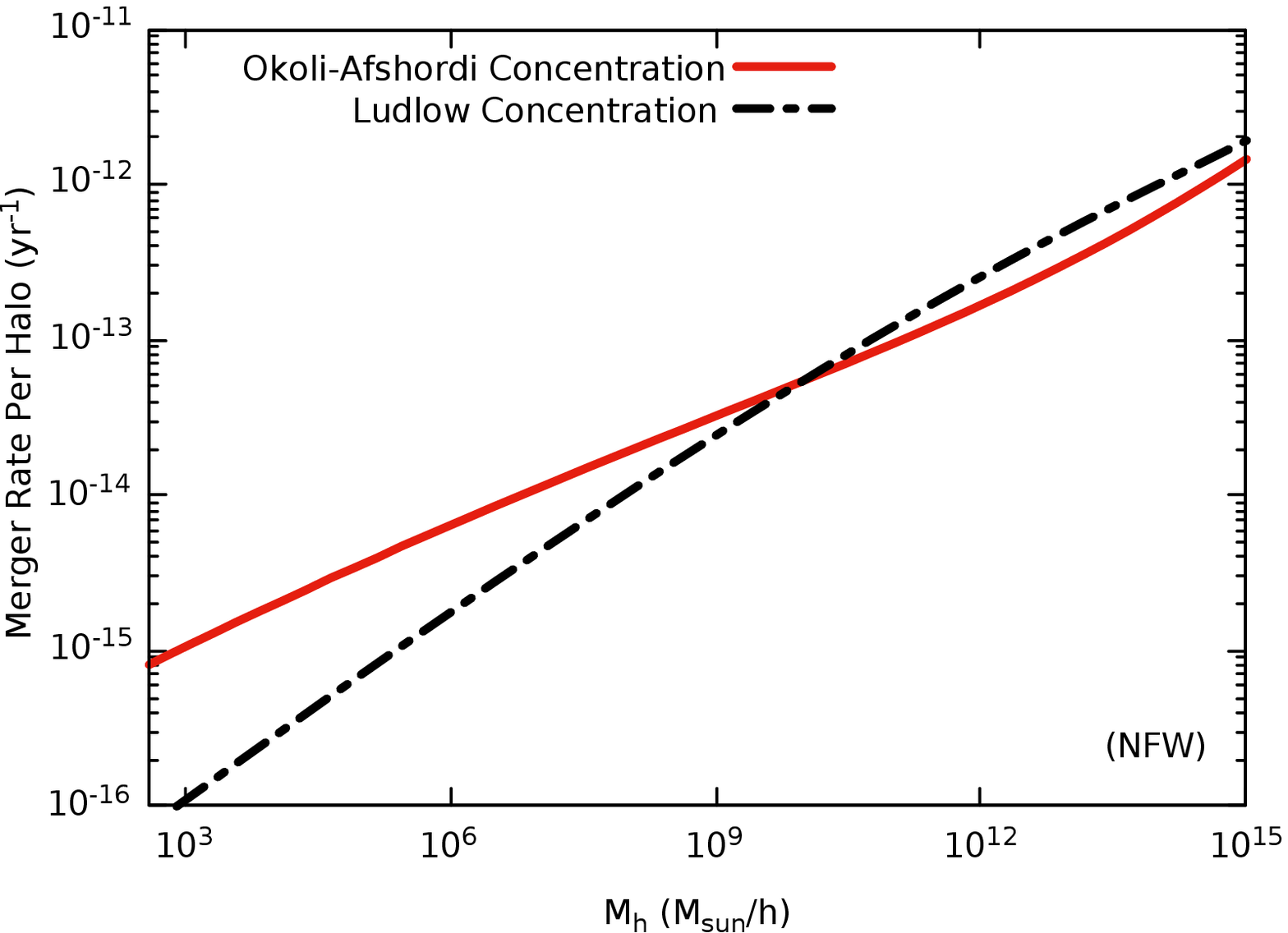}
\\ \hspace*{0.5cm}
\\
\includegraphics[width=1\textwidth]{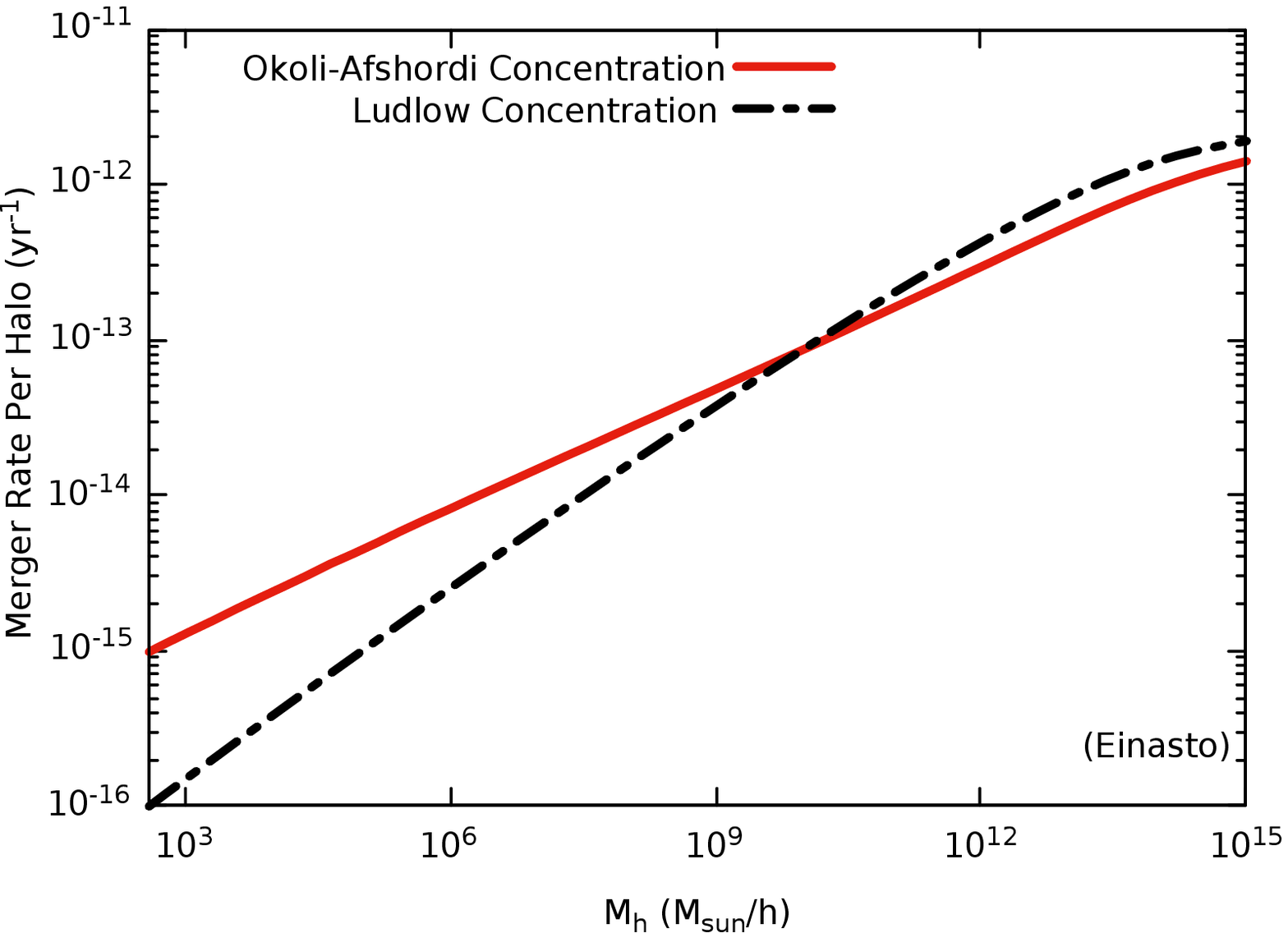}
\caption{(color online) The PBH merger rate in each halo
considered with the NFW profile (top) and the Einasto profile
(bottom). The solid (red) lines represent the merger rate for the
ellipsoidal-collapse model with the O-A concentration-mass, and
the dot-dashed (black) lines show the merger rate for the
spherical-collapse model with the Ludlow concentration-mass
relation.}
 \label{fig:per_year}
\end{minipage}
\end{figure}

To calculate the halo velocity dispersion, one can use its
relation to the maximum velocity in an $r_{\rm max}$ radius, which
has been introduced in Ref.~\cite{prada} as
\begin{equation}
    v_{\rm disp}=\frac{v_{\rm max}}{\sqrt{2}}=\sqrt{\frac{GM(r<r_{\rm max})}{r_{\rm max}}}.
\end{equation}

In this work, we assume that the relative velocity distributions
of PBHs in a halo are random and follow the Maxwell-Boltzmann
statistics. Hence, one can write the velocity probability
distribution function as
\begin{equation}\label{3}
    P(v_{\rm pbh}, v_{\rm disp})=A_{0}\left[\exp\left(-\frac{v_{\rm pbh}^{2}}{v_{\rm disp}^{2}}
    \right)-\exp\left(-\frac{v_{\rm vir}^{2}}{v_{\rm disp}^{2}}\right)\right],
\end{equation}
where $A_{0}$ is determined when the condition
$4\pi\int_{0}^{v_{\rm vir}}P(v)v^{2}dv=1$ is satisfied and a
cutoff is considered at the halo virial velocity.

It is clear from Eqs.~(\ref{13}), (\ref{15}) and (\ref{16}) that,
in order to calculate the merger rate in each halo, the
mass-concentration relation, $C(M_{\rm vir})$, has to be
determined. For this purpose, according to the initial conditions
governing the dark matter halos during the collapse, various
results can be found. In Ref.~\cite{bird}, the merger rate has
been performed by using two famous spherical concentration-mass
relations of Prada, et al.~\cite{prada} and Ludlow, et
al.~\cite{ludlow}.

In this research, we have employed the ellipsoidal-collapse
concentration-mass relation introduced in Ref.~\cite{afshordi}
that we refer to it as Okoli-Afshordi ({\bf O-A})
concentration-mass relation. Furthermore, for the Einasto density
profile, we have chosen the value of the shape parameter presented
in Ref.~\cite{Klypin:2014kpa}. Also, we have set the mass of PBHs
to be $30~M_{\odot}$. In Fig.~\ref{fig:per_year}, we have
indicated the merger rate of PBHs per halo as a function of halo mass
by considering the O-A concentration-mass relation as an
ellipsoidal model, and the Ludlow concentration-mass relation for
a spherical model obtained in Ref.~\cite{bird}. The results show
that the merger rate of PBHs grows with increasing halo mass for both
models. A noteworthy point is that in smaller mass halos for the
case of the O-A model the merger rate is almost one order of
magnitude larger than the corresponding result for the Ludlow model.

In the following, we propose to determine the effect of these
changes on the total merger rate of PBHs in a given volume and
time interval.
\begin{table*}[t]
  \caption{General information on the PBH total merger rate per
  unit time and unit comoving volume for the ellipsoidal
  and spherical models in terms of two density profiles of the
  NFW and the Einasto at the present-time universe. The
  PBH mass has been considered to be $30~M_{\odot}$.}
    \centering
    \begin{tabular}{c|c|c|c|c}
        \hline
        \hline
        Halo Density Profile & Halo Mass Function & $\rm C(M)$ &
        Lower Limit Halo Mass $\rm (M_{\odot})$& Total Merger Rate $(\rm Gpc^{-3}\rm yr^{-1})$\\ [0.5ex]
        \hline
                & P-S & Ludlow & $4\times 10^{2}$ & $1.40$ \\
                & S-T & O-A & $4\times 10^{2}$ & $15.06$ \\
                & P-S & Ludlow & $4\times 10^{3}$ & $0.44$ \\
        NFW & S-T & O-A & $4\times 10^{3}$ & $3.29$ \\
                & P-S & Ludlow & $4\times 10^{4}$ & $0.14$ \\
                & S-T & O-A & $4\times 10^{4}$ & $0.69$ \\
        \hline
                & P-S & Ludlow & $4\times 10^{2}$ & $1.93$ \\
                & S-T & O-A & $4\times 10^{2}$ & $24.03$ \\
                & P-S & Ludlow & $4\times 10^{3}$ & $0.61$ \\
        Einasto & S-T & O-A & $4\times 10^{3}$ & $5.62$ \\
                & P-S & Ludlow & $4\times 10^{4}$ & $0.20$ \\
                & S-T & O-A & $4\times 10^{4}$ & $1.31$ \\
        \hline
        \hline
    \end{tabular}
    \label{table1}
\end{table*}
\subsection{Total Merger Rate}
\subsubsection{Present-Time Universe}
Up to here, the merger rate has been considered within each dark
matter halo. However, as the gravitational wave detectors
statistically receive the cumulative events, it is necessary to
calculate the total merger event rate per unit volume and per unit
time. For this purpose, convolving the merger rate per halo,
$\Phi(M_{\rm h})$, with the halo mass function, $dn/dM_{\rm h}$,
leads to the total merger event rate per unit volume and per unit
time as
\begin{equation}\label{tot_mer}
    \mathcal{R}=\int\frac{dn}{dM_{\rm h}}\Phi(M_{\rm h})dM_{\rm h},
\end{equation}
where $M_{\rm h}$ is the halo mass, which can be estimated as the
virialized mass, $M_{\rm vir}$. As is clear from Eqs.~(\ref{mf}),
(\ref{mf1}) and (\ref{mf2}), the exponential decay of the mass
function means that the upper limit of the integral does not
affect the final result. Instead, the lower limit plays a crucial
role.

\begin{figure}[t!]
\begin{minipage}{1\linewidth}
\includegraphics[width=1\textwidth]{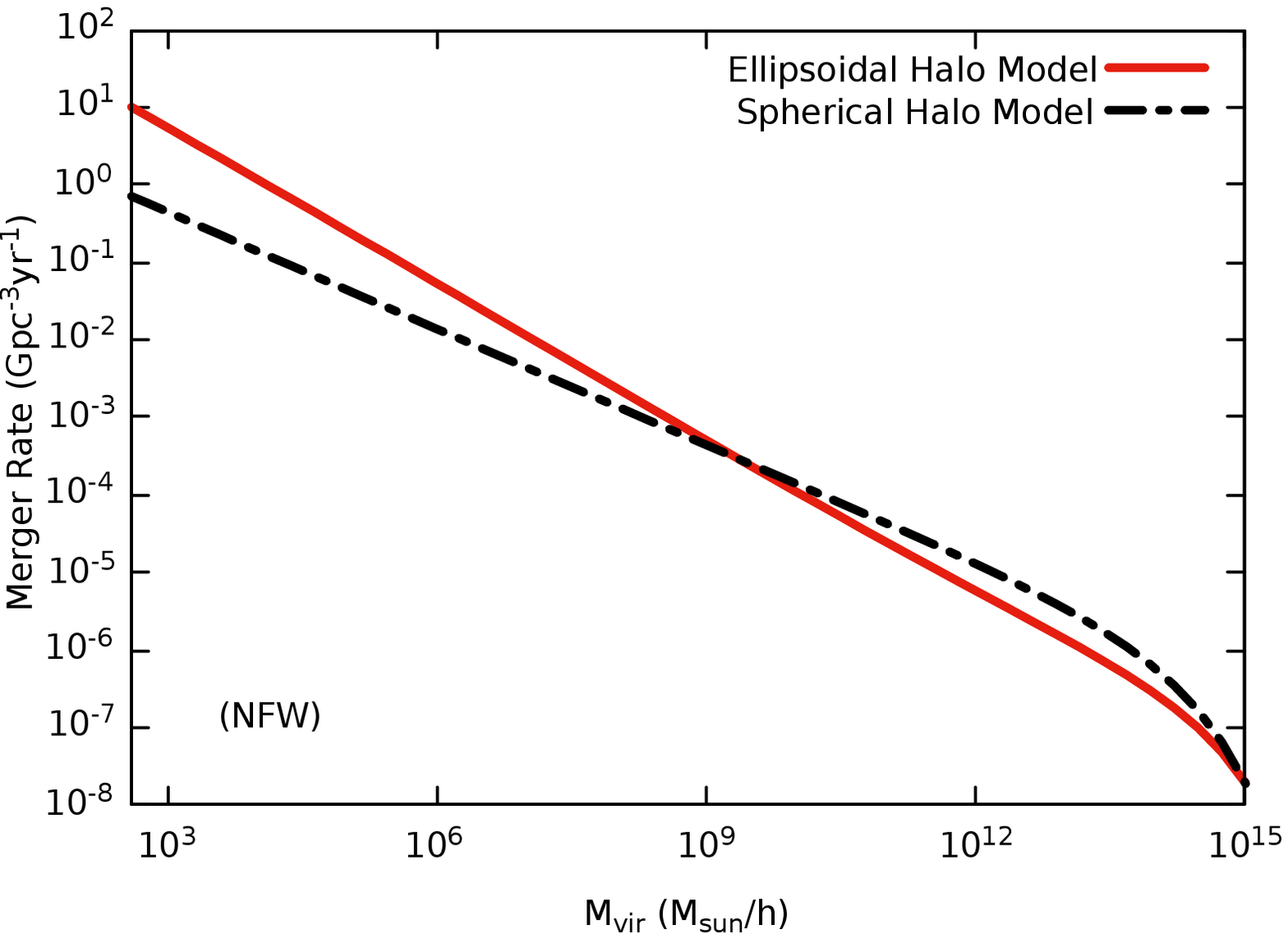}
\\ \hspace*{0.5cm}
\\
\includegraphics[width=1\textwidth]{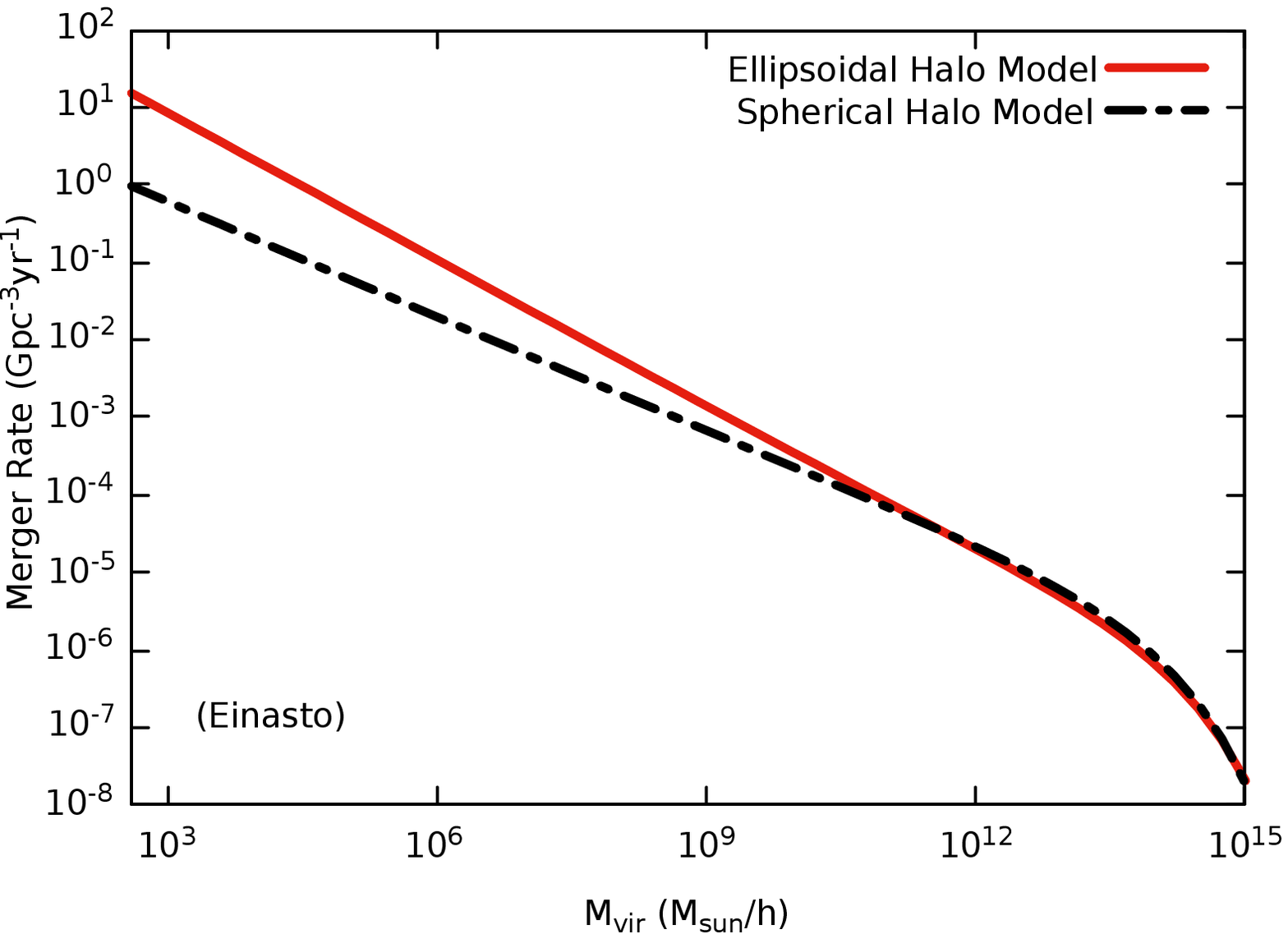}
\caption{(color online) The PBH merger event rate per unit volume
and per unit time for the spherical- and ellipsoidal-collapse
models with the NFW profile (top) and the Einasto profile
(bottom). The solid (red) lines represent the ellipsoidal halo
model with the S-T mass function and the O-A concentration-mass
relation, and the dot-dashed (black) lines show the spherical halo
model with the P-S mass function and the Ludlow concentration-mass
relation.} \label{fig:per_year_gpc}
\end{minipage}
\end{figure}

It should be noted that the merger time of BH binaries is a
function of the velocity dispersion of halos (from hours to
kiloyears)~\cite{OLeary:2008myb}. Thus, BH binaries, that are
formed due to dissipative two-body encounters, have merger time
much shorter than the age of the universe.  Moreover,
non-dissipative three-body encounters can also lead to the
formation of BH binaries. These types of binaries have no strong
enough binding energy to decay instantaneously via the emission of
gravitational radiation. Therefore, these binaries frequently lead
to the formation of wide binaries in a way that their merger times
are longer than a Hubble time~\cite{quinlan}. Consequently, these
binaries should~not affect the aLIGO merger rate.

It is known that the smallest halos are the most concentrated
halos, and those have already become virialized.  Moreover, the
dynamical relaxation processes lead to the evaporation of the
smallest halos through the ejection of objects. The timescale of
such an evaporation has been stated to be $t_{\rm
ev}=-N(dN/dt)^{-1}\propto t_{\rm th}$, where $N=M_{\rm h}/M_{\rm
pbh}$ is the number of BHs in a halo, $dN/dt$ is the ejection rate
of objects, and $t_{\rm th}$ is the half-mass relaxation
time~\cite{binney2011galactic}. The calculated evaporation time
for halos with a typical mass $400~M_{\odot}$ (while containing
PBHs each with mass of $30~M_{\odot}$) is about $3~\rm
Gyr$~\cite{bird}. However, it would be plausible that halos with
masses less than $400~M_{\odot}$ (while maintaining the same PBH
mass) have the evaporation timescale less than $3~\rm Gyr$.

On the other hand, during the matter-dominated era, the halo
evaporation is a process that is compensated by the accretion of
outer objects onto the halo and/or the formation of new halos due
to the merging of smaller objects. While the relative separation
of structures increases during the dark energy dominated era (i.e.
recent $3~\rm Gyr$), due to the increasing expansion rate of the
universe, the compensatory factors (i.e. merging and accretion)
against the halo evaporation become very slow. Hence, it can be
inferred that the signal from halos with masses $M_{\rm
h}<400~M_{\odot}$ would be negligible.

To quantify the total merger rate introduced in
Eq.~(\ref{tot_mer}), two crucial quantities, namely the halo mass
function and the concentration-mass relation, must be specified in
proportion to the dark matter halo formation conditions. The idea
is to look at the merger rate of PBHs for the ellipsoidal-collapse
halo models. For this purpose, we use the S-T mass function and
the O-A concentration-mass relation which have been introduced for
the ellipsoidal-collapse halo models.

Fig.~\ref{fig:per_year_gpc} shows the merger rate of PBHs for the
ellipsoidal halo models per unit time and per unit volume, and
compares it with the results of the spherical model, which has
been evaluated in Ref.~\cite{bird}, while taking into account the
NFW density profile (top) and the Einasto density profile
(bottom). In the ellipsoidal model, the S-T mass function and the
ellipsoidal O-A concentration-mass relation have been considered,
while in the spherical model, the P-S mass function and the Ludlow
concentration-mass relation are used. As expected, the total
merger rate of PBHs for ellipsoidal models, like the spherical
models, increases with decreasing halo mass due to the
significance of merger events in the smallest halos. For the halo
masses larger than $M_{h}>(10^{9} - 10^{10}) M_{\odot}$, the
merger rate is approximately the same in both models. However, for
masses smaller than $M_{h}<(10^{9} - 10^{10}) M_{\odot}$ with the
ellipsoidal model, it is prominently increased by about one order
of magnitude compared with the spherical model. The total merger
rate has been obtained by integrating over the surface below the
curves, and the results for different lower limits of halo masses
have been presented in Table~\ref{table1} for the present-time
universe. This table, and also Fig.~\ref{fig:per_year_gpc},
indicates that the PBH total merger rate decreases as the lower
limit of the halo mass increases.

\subsubsection{Redshift Evolution of PBH Merger Rate}
The time evolution of the BH merger rate has always been one of
the most interesting topics to study. Because it can potentially
provide a clearer picture to distinguish among the BH formation
scenarios~\cite{Sasaki:2018dmp}. On the other side, the
sensitivity of the aLIGO detectors can observe the binaries up to
$z\sim 0.75$ which includes a comoving volume around $50~\rm
Gpc^{3}$~\cite{Aasi:2013wya,TheLIGOScientific:2016zmo,LIGOScientific:2018jsj}.

Here, we intend to represent the redshift evolution of the PBH
merger rate for the ellipsoidal and spherical halo models, and
compare their results with each other. In this respect, by
definition of the concentration parameter (Eq. (\ref{conc})), it
is obvious that $C$ is redshift dependent through the halo virial
radius~\cite{Mandic:2016lcn}. On the other hand, the halo mass
function also depends on the redshift through $\sigma(M, z)$.
Thus, via Eq.~(\ref{tot_mer}), one can obtain the redshift
evolution of the PBH total merger rate $\mathcal{R}(z)$.

\begin{figure}[t!]
\begin{minipage}{1\linewidth}
\includegraphics[width=1\textwidth]{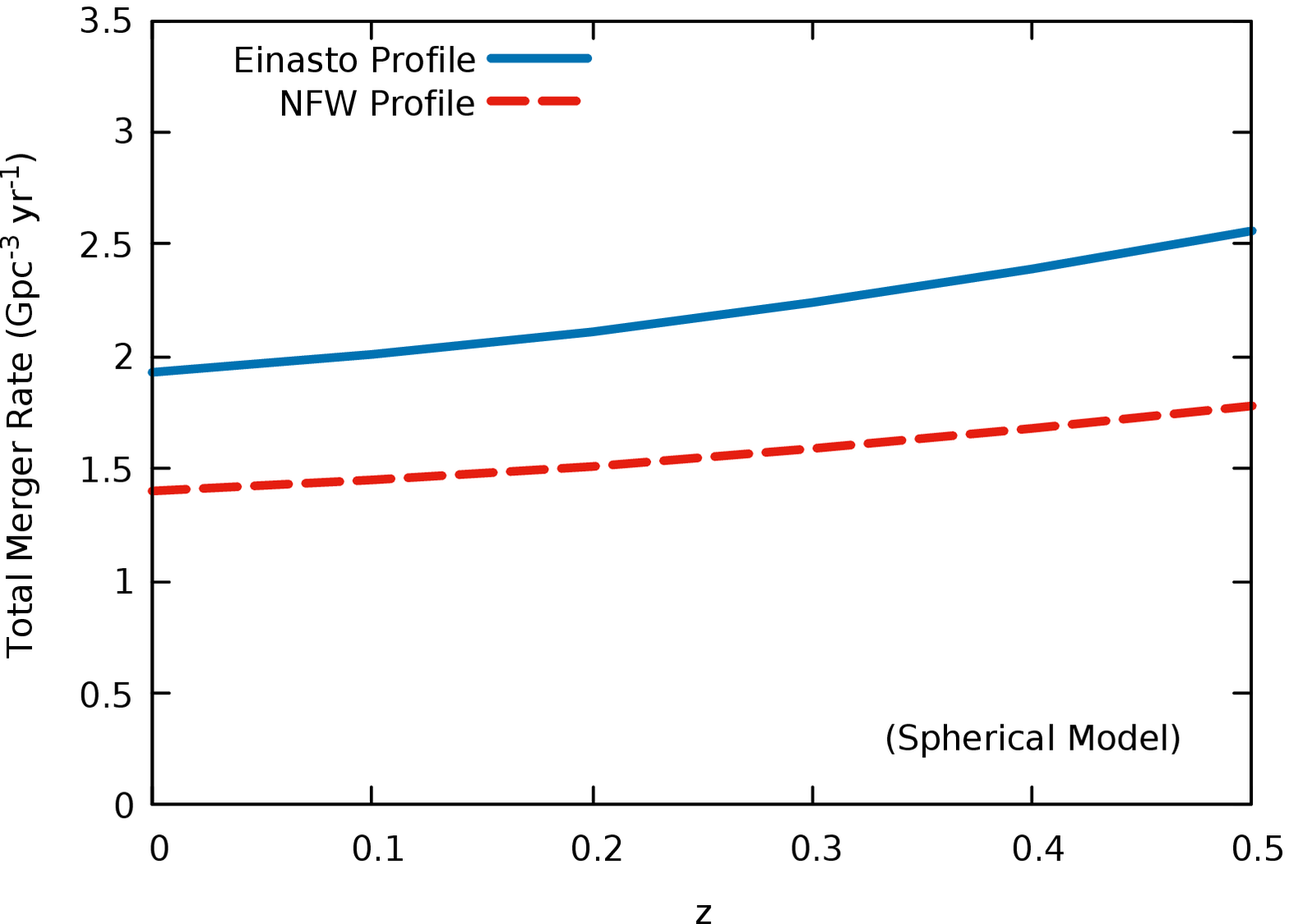}
\\ \hspace*{0.5cm}
\\
\includegraphics[width=1\textwidth]{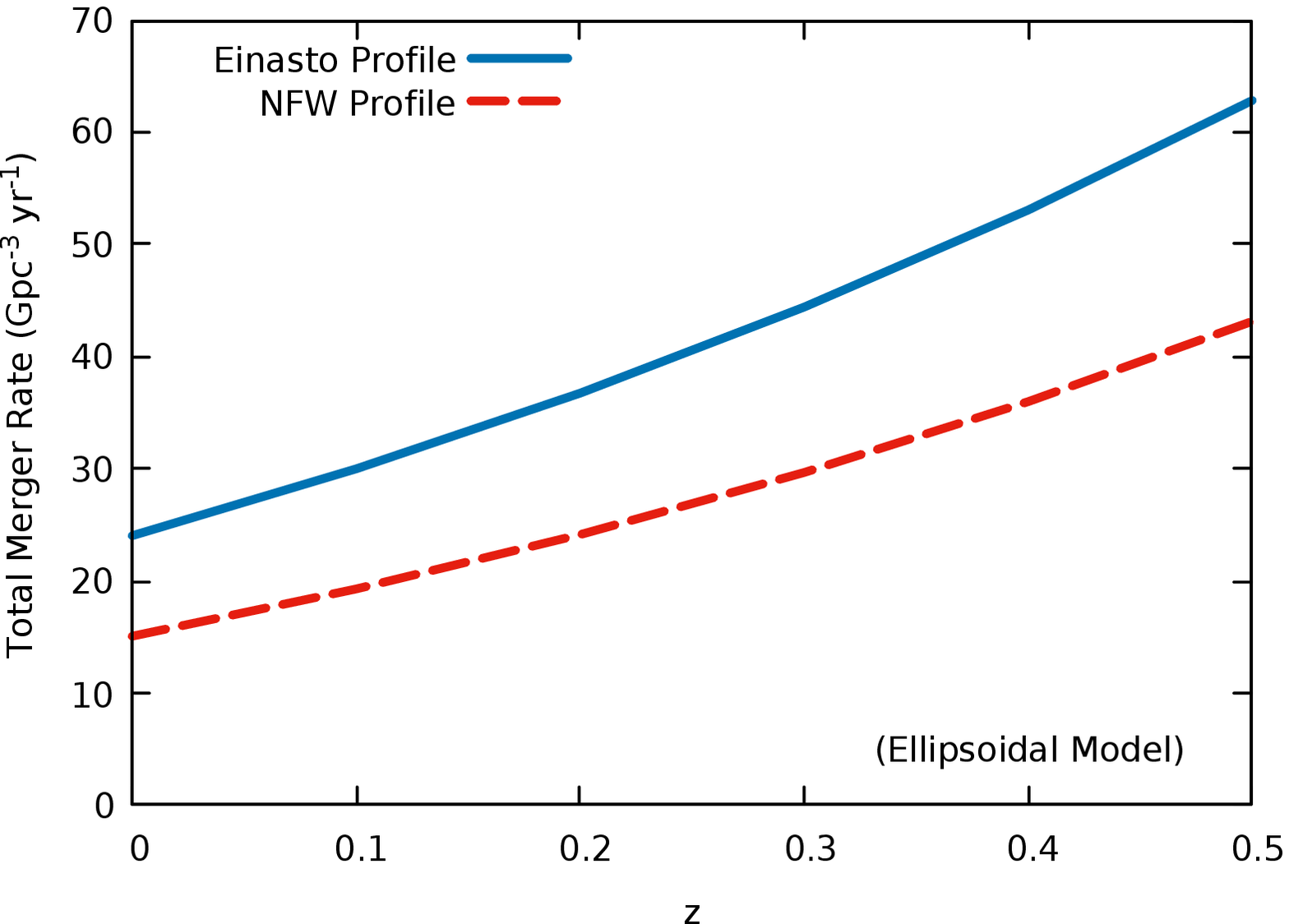}
\caption{(color online) The PBH total merger event rate per unit
source time and unit comoving volume for the spherical (top) and
the ellipsoidal (bottom) collapse models as a function of
redshift. The solid (blue) lines indicate the calculations
considered the Einasto density profile, and the dashed (red) lines
are for the NFW density profile.} \label{fig:per_redshift}
\end{minipage}
\end{figure}

In Fig.~\ref{fig:per_redshift}, we have demonstrated the total
merger event rate for both the ellipsoidal and spherical models as
a function of redshift, wherein two halo profile models of the NFW
and the Einasto have been compared. The results indicate that the
merger rate in higher redshifts has been more significant than the
present-day universe, which is consistent with the results
obtained in Refs.~\cite{Sasaki:2018dmp,Gow:2019pok}. Furthermore,
this figure indicates that the evolution of the PBH merger rate in
the ellipsoidal model is much more pronounced than in the
spherical model. Hence, the results completely confirm what we
have expected, since the ellipsoidal model uses the S-T mass
function, which is sensitive to redshift changes, while the
spherical model is not significantly sensitive to redshift changes
due to using the P-S mass function.

\subsubsection{Constraint on PBH Fraction}
As the last part, let us concentrate on the expected PBH fraction,
$f_{\rm pbh}$, extracted from the ellipsoidal-collapse halo model.
The problem of PBH abundance has been an important issue since the
beginning of the emergence of the PBH scenario. Moreover, one of
the most important constraints imposed on PBHs is their abundance
in the late-time universe. The fraction of PBHs determines their
contribution to dark matter. Many studies have been performed in
this area, and today it is believed that this fraction is lower
than one for most of the PBH mass
ranges~\cite{Sasaki:2016jop,Raidal:2017mfl,Ali-Haimoud:2017rtz,Kocsis:2017yty,Wang:2016ana,Kohri:2018qtx,Wang:2019kaf,Carr:2020gox,
DeLuca:2020fpg, DeLuca:2020qqa, Wong:2020yig}. It means that dark
matter consists of several components, one of which is the PBHs.
However, it should be noted that a small mass range of PBHs, known
as asteroid-mass
PBHs~\cite{Carr:2020gox,Carr:2020xqk,Katz:2018zrn,Smyth:2019whb,Montero-Camacho:2019jte,coogan,Picker:2021jxl,Ray:2021mxu},
has~not yet been strongly constrained, and this window (i.e.
$10^{-17}~M_{\odot}\leqslant M_{\rm pbh}\leqslant
10^{-12}~M_{\odot}$) may make up a significant fraction of dark
matter.

\begin{table*}[t]
    \caption{The PBH total merger rate per unit time and
    unit comoving volume for different masses of PBHs, i.e. $M_{\rm pbh} = 10,20,50$ and $100~M_{\odot}$,
    while considering the ellipsoidal halo model in terms of the NFW and the
    Einasto density profiles at the present-time universe.}
    \centering
    \begin{tabular}{c|c|c}
        \hline
        \hline
        PBH Mass $(M_{\odot})$ & Density Profile &Total Merger Rate $(\rm Gpc^{-3}\rm yr^{-1})$\\ [0.5ex]
        \hline
        10 & NFW & $33.41$ \\
        10 & Einasto & $51.25$ \\
       \hline
        20 & NFW & $22.38$ \\
        20 & Einasto & $35.01$ \\
        \hline
        50 & NFW & $11.49$ \\
        50 & Einasto & $18.60$ \\
        \hline
        100 & NFW & $8.18$ \\
        100 & Einasto & $13.48$ \\
        \hline
        \hline
    \end{tabular}
    \label{table2}
\end{table*}

On the other hand, one of the best references to investigate
suitable models of dark matter halos is to compare the obtained
merger rate from each model with the determined one via the aLIGO
detectors. However, since the aLIGO mergers could contain the
astrophysical BH mergers, the inferred PBH fraction from
theoretical models is an upper limit that is allowed by the aLIGO
observations.

\begin{figure}[t!]
\begin{minipage}{1\linewidth}
\includegraphics[width=1\textwidth]{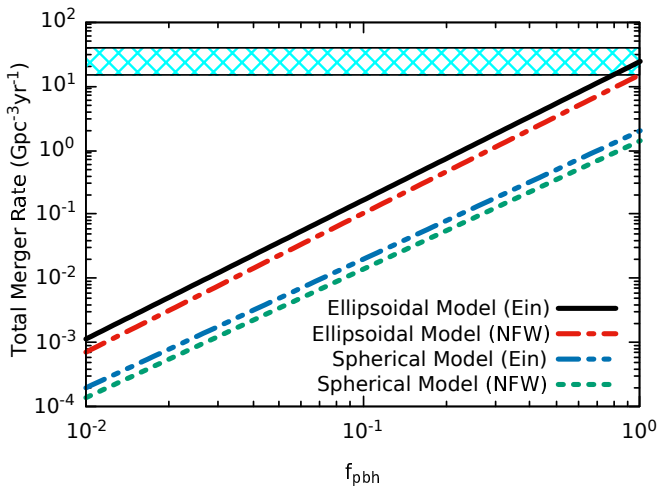}
\caption{(color online) The PBH total merger event rate for the
ellipsoidal and the spherical models with respect to the fraction
of PBHs, $f_{\rm pbh}$. The solid (black) line indicates the total
merger rate for the ellipsoidal model with the S-T mass function
and the Einasto density profile, while the dot-dashed (red) line
shows the related result for the NFW density profile. The
dot-dot-dashed (blue) line represents the total merger rate for
the spherical model with the P-S mass function and the Einasto
density profile, and the dotted (green) line shows the
corresponding result for the NFW density profile. The shaded
(cyan) band is the estimated merger rate from the third observing
run (O3) of the aLIGO detectors, i.e. $(15.3-38.8)~\rm
Gpc^{-3}yr^{-1 }$.} \label{fig:per_abound}
\end{minipage}
\end{figure}

In Fig.~\ref{fig:per_abound}, we have depicted the PBH total
merger rate as a function of $f_{\rm pbh}$ for the ellipsoidal
halo model while considering those two density profiles of the NFW
and the Einasto, and have compared the results with the
corresponding ones for the spherical model~\cite{Sasaki:2018dmp}.
In this figure, the shaded band is the estimated merger event rate
$(15.3-38.8)~\rm Gpc^{-3}yr^{-1 }$ by the third observing run (O3)
of the aLIGO detectors~\cite{Abbott:2020niy}. Although there are
some theoretical uncertainties, the results indicate that the
calculated merger rate, for the $30~M_{\odot}$ PBHs while
considering the ellipsoidal halo model, falls within the aLIGO
window when $f_{\rm pbh}\simeq 1$. However, the corresponding
result of the spherical halo model does~not enter in this window
for any value of $f_{\rm pbh}\leqslant 1$. This means that
ellipsoidal halo models are capable to generate enough PBH mergers
consistent with the estimated ones by the aLIGO detectors.
Whereas, if the spherical models are trusted, the population of
PBHs should be reduced by some mechanism and most of the aLIGO
mergers must be of astrophysical origins. Moreover, as mentioned
earlier, the aLIGO detectors can probe events up to $z\sim 0.75$
that corresponds to the comoving volume around $50~\rm Gpc^{3}$.
Thus, within the context of ellipsoidal halo models, over one
year, one expects that aLIGO can detect approximately $(750-1200)$
events with $f_{\rm pbh}\simeq 1$, while the number of events
during this time should be at least $(10-15)$ if $f_{\rm
pbh}>10^{-1}$, and at least one if $f_{\rm pbh}>10^{-2}$.

Up to here, we have have considered the PBHs only with
$(30~M_{\odot})$ masses, and one may ponder whether the results
change with smaller or larger masses. In these cases, recall the
argument regarding limits on the lower bound of halo masses that
have~not yet evaporated by the present-time. Then, for PBHs with
masses less than $30~M_{\odot}$ resided in dark matter halos, it
can be inferred that the smallest host halos, which have an
evaporation time around $3~\rm Gyr$, should have masses less than
$400~M_{\odot}$. Similarly, the smallest host halos containing
PBHs with masses larger than $30~M_{\odot}$ must have masses
greater than $400~M_{\odot}$. In addition, the PBH total merger
rate depends sensitively on the lower limit of the halo mass in a
way that the smaller the halo mass, the higher the total merger
rate. Nevertheless, to achieve the merger rate of PBHs with
different masses, one should consider this criterion for the lower
limit of halo masses, and repeat the calculations.

\begin{figure}[t!]
\begin{minipage}{1\linewidth}
\includegraphics[width=0.95\textwidth]{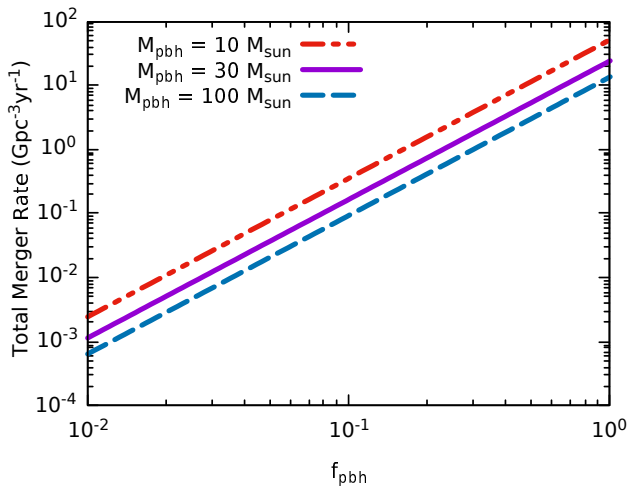}
\caption{(color online) The PBH total merger event rate for the
ellipsoidal model as a function of the PBH fraction and mass. The
dot-dot-dashed (red) line, the solid (purple) line, and the dashed
(blue) line exhibit this dependency for $M_{\rm pbh} = 10, 30$ and
$100~M_{\odot}$, respectively. In here, the Einasto density
profile has been considered.} \label{fig:fraction_event}
\end{minipage}
\end{figure}

In this regard, in Fig.~\ref{fig:fraction_event}, we have depicted
the total merger rate of PBHs for ellipsoidal models in terms of
the PBH fraction and mass. The results have been presented for
several masses of PBHs (i.e. $M_{\rm pbh}=10, 30$ and
$100~M_{\odot}$). As it is clear, the total merger rate changes
inversely with the PBH mass. Also, the merger rate of smaller
masses, compared to larger ones, falls within a wider range of the
aLIGO band. It should be noted that we have only shown the results
for the Einasto density profile. The total merger rate of PBHs
with different masses in ellipsoidal halo models has been
presented in Table~\ref{table2} for the NFW and the Einasto
density profiles.

\begin{figure}[t!]
\begin{minipage}{1\linewidth}
\includegraphics[width=0.94\textwidth]{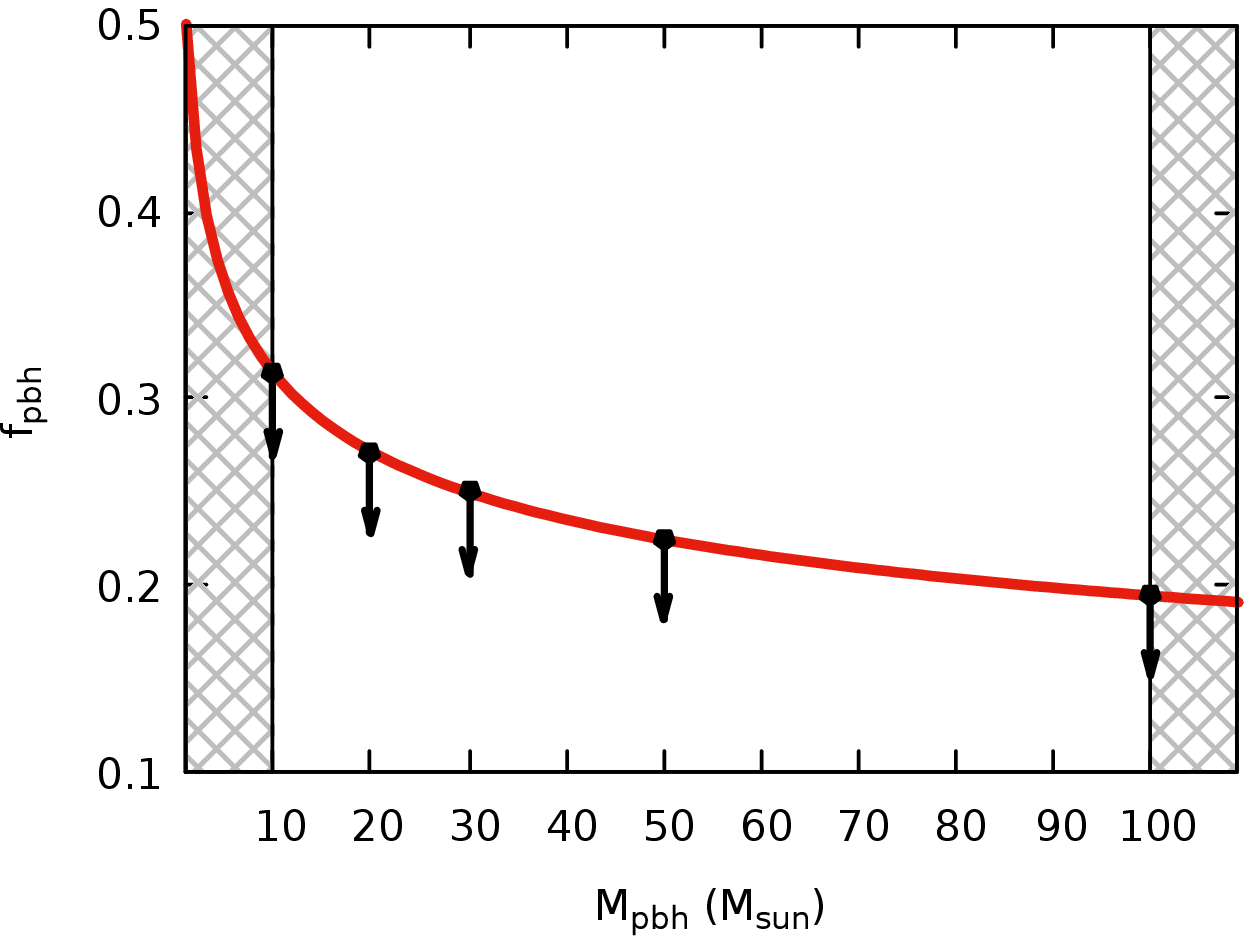}\\ \hspace{-0.5cm}
\\ \hspace{-0.5cm}
\includegraphics[width=0.98\textwidth]{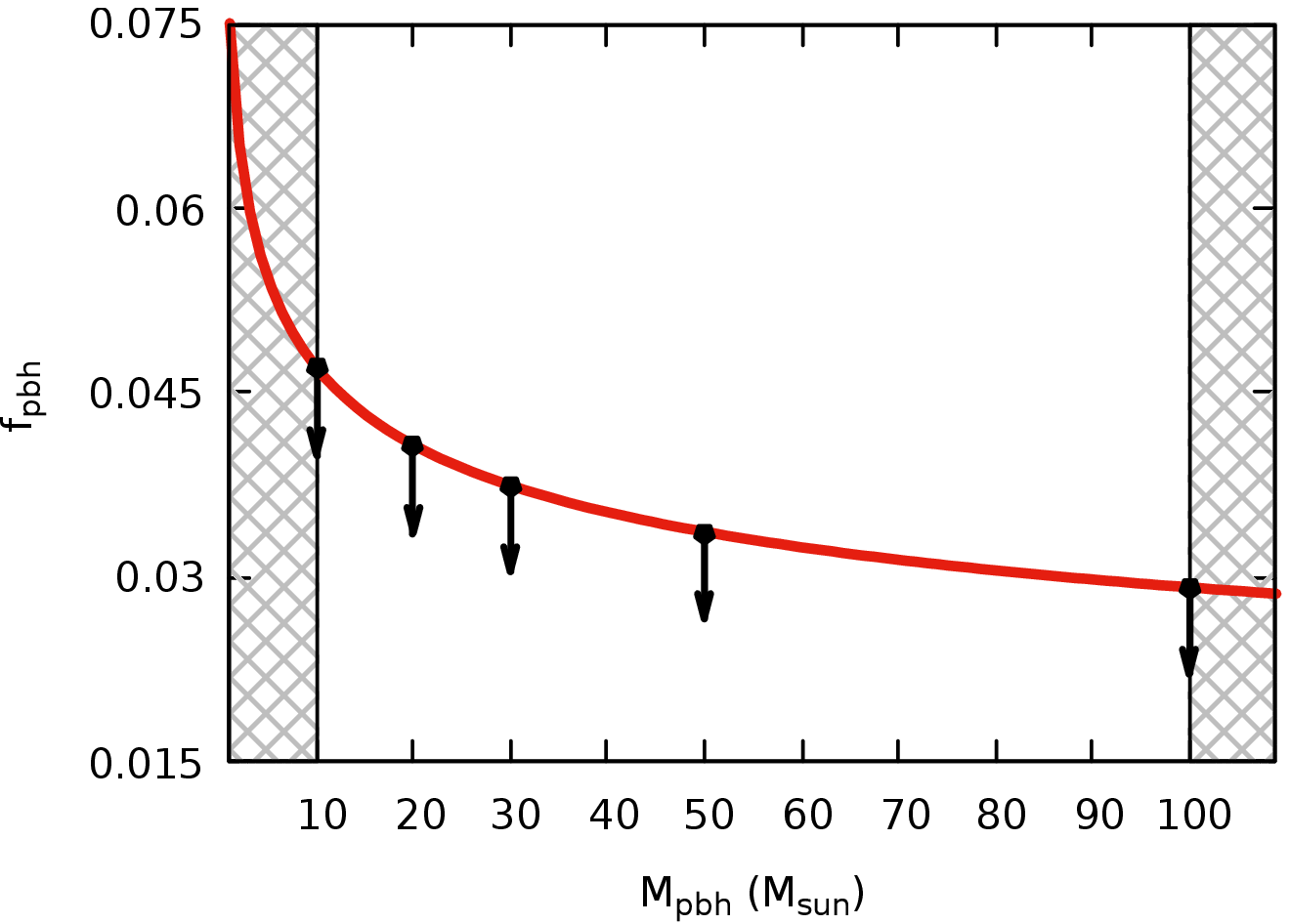}
\caption{(color online) The expected upper bounds on the fraction
of PBHs, $f_{\rm pbh}$, as a function of their masses in the rage
$10~M_{\odot}<M_{\rm pbh}<100~M_{\odot}$, while considering the
ellipsoidal halo model. The top plot has been calibrated for the
situation in which one should expect to detect at least one
$(30~M_{\odot}-30~M_{\odot})$ event in the comoving volume $1~\rm
Gpc^{3}$ and over one year, while the bottom plot has been
quantitated for the same event, but in the comoving volume $50~\rm
Gpc^{3}$.} \label{fig:fraction_mass}
\end{minipage}
\end{figure}

Furthermore, it is constructive to discuss about the relation
between the fraction of PBHs and their masses. According to the
relations obtained for the total merger rate in
Refs.~\cite{bird,Ali-Haimoud:2017rtz}, one can estimate the
dependence of the fraction of PBHs on their masses to be $f_{\rm
pbh}\sim\left(M_{\rm pbh}/30~M_{\odot}\right)^{-11/53}$. This
relation is consistent with our findings on the merger rate and
the constraints of PBHs with different masses. In
Fig.~\ref{fig:fraction_mass}, we have depicted the expected upper
bounds on the fraction of PBHs in terms of their masses for
ellipsoidal halo models while considering a narrow mass
distribution for PBHs, i.e. $10~M_{\odot}<M_{\rm
pbh}<100~M_{\odot}$. It is clear that the fraction of PBHs
decreases as their masses increase. We have also marked upper
limits on the fraction of PBHs with smaller and larger masses than
$30~M_{\odot}$. The top plot shows this relation for the situation
in which one should expect to detect at least one
$(30~M_{\odot}-30~M_{\odot})$ event over one year in the comoving
volume $1~\rm Gpc^{3}$, whereas the bottom plot indicates the
corresponding result for the same event but in the comoving volume
$50~\rm Gpc^{3}$.

\section{Conclusions}\label{sec. iv}

In this work, we have focused on the modeling of the stellar-mass
PBH merger rate assuming the ellipsoidal-collapse of dark matter
halos. Specifically, to perform this task, we have considered two
crucial components that have been calculated for the case of
ellipsoidal-collapse dark matter halos, namely the S-T mass
function, and the ellipsoidal concentration-mass relation obtained
in Ref.~\cite{afshordi}. The main idea behind the extraction of
these two important components in the ellipsoidal-collapse halos
has been to propose a dynamical threshold overdensity,
$\delta_{\rm ec (\nu)}$, instead of a constant threshold one,
$\delta_{\rm sc}=1.686$, which has already been introduced for the
spherical-collapse dark matter halos. This generalization of
threshold overdensity has led to a more realistic model that fits
the observational data.

Subsequently, we have mentioned the scattering amplitude of the
PBHs by considering the encounter conditions in the medium of dark
matter halos. We have also used the NFW and the Einasto density
profiles. Under these assumptions, we have calculated the merger
rate of PBHs per halo for the ellipsoidal-collapse halos and have
compared it with the corresponding result of the
spherical-collapse ones. It has been observed that, in the
smallest halo masses, the merger rate per halo for the ellipsoidal
model is about one order of magnitude larger than that in the
corresponding one for the spherical model. The results amplify the
fact that the probability of binary black hole formation within
halos with the lowest mass is more prominent because these halos
are more compact and have less the virial velocity compared with
the larger mass halos.

Furthermore, we have focused on the PBH merger rate per unit
volume and per unit time in an ellipsoidal halo model. In these
calculations, the significance of the merger rate in the
ellipsoidal model has been evident compared with the results of
the spherical model for the halos with the lowest mass. As a
result, cumulatively, the significance of the merger rate in the
ellipsoidal halos is confirmed.

Given the possibility of the PBH binary formation during the age
of the universe, as an interesting case study, we have calculated
their total merger rate evolution as a function of redshift. It
has been observed that the evolution of the total merger rate of
PBHs in the case of ellipsoidal halo models with redshift is more
sensitive than its evolution in the spherical model. This
sensitivity is due to the consideration of the dynamical threshold
overdensity in the ellipsoidal model.

We have plotted the total merger rate of $(30~M_{\odot})$ PBHs in
the ellipsoidal halo models in terms of their fraction, and have
compared it with the corresponding results of the spherical halo
models obtained in Ref.~\cite{bird}. The criterion used for this
comparison is the merger rate estimated by the aLIGO detectors
during the third observing run (O3), i.e. $(15.3-38.8)~\rm
Gpc^{-3}yr^{-1}$. Such an evaluation is important because it can
estimate the contribution of PBHs in dark matter. We have shown
that the total merger rate of $(30~M_{\odot})$ PBHs in the
ellipsoidal halo models, despite some theoretical uncertainties,
falls within the aLIGO window, while the corresponding results for
the spherical halo models are~not consistent with the aLIGO merger
rate. This result suggests that the merger of $(30~M_{\odot})$
PBHs within the framework of the ellipsoidal halo models is still
a potential candidate for dark matter, but the spherical halo
models are no~longer able to justify the PBH mergers compared to
the aLIGO window. Otherwise, it reinforces the argument that the
aLIGO detectors are more likely to detect the BHs of astrophysical
origin if the spherical halo models would be reliable. Besides,
based on the result of the ellipsoidal halo models for
$(30~M_{\odot})$ PBHs, it can also be inferred that over one year
aLIGO can detect approximately $(750-1200)$ events with $f_{\rm
pbh}\simeq 1$, while the number of events during this time should
be at least $(10-15)$ if $f_{\rm pbh}>10^{-1}$, and at least one
if $f_{\rm pbh}>10^{-2}$. Relying on the evaporation time of the
smallest halos and repeating the calculations, we have indicated
that the total merger rate for the ellipsoidal halo models in
terms of the PBH fraction for several different masses of PBHs.
The results suggest that the total merger rate changes inversely
with the PBH mass. Also, we have estimated the relation between
the fraction of PBHs and their masses, and have shown it for a
mass distribution between $(10-100)~M_{\odot}$. The results
indicate that the fraction of PBHs for the ellipsoidal halo models
should be around $f_{\rm pbh}\sim {\cal O}(10^{-1})$, if one could
expect to detect at least one $(30~M_{\odot}-30~M_{\odot})$ PBH
event in the comoving volume $1~\rm Gpc^{3}$ over one year.
Whereas the corresponding fraction should be around $f_{\rm
pbh}\sim {\cal O}(10^{-2})$ for the same event but in the comoving
volume $50~\rm Gpc^{3}$. Interestingly, the constraint obtained in
this work can be potentially stronger than the one
obtained~\cite{bird} from the merger rate of PBHs in the spherical
halo models.

However, within the mentioned mass range of PBHs, there are other
strong observational constraints that come from the accretion
limits from the observed number of X-ray
binaries~\cite{Inoue:2017csr}, the Planck data on the CMB
anisotropies~\cite{Ali-Haimoud:2016mbv,Serpico:2020ehh}, dynamical
processes from star clusters in nearby dwarf
galaxies~\cite{Brandt:2016aco,Koushiappas:2017chw} and the
gravitational lensing of type Ia
supernovae~\cite{Zumalacarregui:2017qqd}.

It is noteworthy to mention that the constraint of PBHs is subject
to many uncertainties including various models of halo, e.g.
spherical and non-spherical collapses, various physical processes
that may lead to the growth (e.g. merging and accretion) or the
evaporation (e.g. the substantial spin) of PBHs, the mass
distribution of PBHs in the dark matter halos, the uncertainties
in the estimated merger rate by the aLIGO detectors, the
considered PBH mass (or the mass range) and contribution of
(astrophysical and primordial) BH binary mergers to the merger
rate recorded by the aLIGO detectors. These uncertainties lead to
the advent of unknown factors in calculations. However, we may
better understand the physics governing those to reach stronger
constraints on PBHs in the future.

\section*{Acknowledgments}
Fakhry and Farhoudi thank the Research Council of Shahid Beheshti
University. The authors gratefully acknowledge the anonymous
referee for the constructive comments.

\end{document}